\documentclass[draftclsnofoot,onecolumn]{IEEEtran}
\IEEEoverridecommandlockouts
\usepackage{hyperref}       
\usepackage{url}            
\usepackage{booktabs}       
\usepackage{amsfonts}       

\usepackage{wrapfig}        
\usepackage[square,sort,comma,numbers]{natbib}
\usepackage{pifont}        

\usepackage{nicefrac}       
\usepackage{microtype}      
\usepackage{bm}
\usepackage{graphicx}
\usepackage{amsmath}
\usepackage{amssymb}
\usepackage{amsfonts}
\usepackage[font=normalsize]{caption}
\usepackage{textcomp}

\usepackage{subcaption}
\usepackage{algorithm}
\usepackage{algpseudocode}
\usepackage{booktabs}
\usepackage{makecell}
\usepackage{colortbl}
\usepackage{enumitem}
\usepackage[english]{babel} 
\definecolor{Gray}{gray}{0.9}
\definecolor{LightCyan}{rgb}{0.75,1,1}
\newcolumntype{y}{>{\columncolor{LightCyan}}c}
\usepackage[table]{xcolor}      
\definecolor{softlavender}{RGB}{233,238,250}  
\usepackage{tabularx}      
\usepackage{longtable} 

\usepackage{booktabs}
\usepackage{arydshln}
\usepackage{multirow}
\usepackage{longtable}

\usepackage[many]{tcolorbox}  
\newtcolorbox{boxL}{
    fontupper = \color{black},
    rounded corners,
    arc = 6pt,
    colframe = black!50, 
    boxrule = 0pt, 
    bottomrule = 4.5pt ,
    breakable,
}
\begin{document}

\title{Repair-R1: Better Test Before Repair }
\author{\IEEEauthorblockN{Haichuan Hu, Alibaba Cloud, 522022320050@smail.nju.edu.cn\\}
\and
\IEEEauthorblockN{Xiaochen Xie, Zhejiang University, xcxie@zju.edu.cn \\ }
\IEEEauthorblockA{Quanjun Zhang, Nanjing University of Science and Technology, quanjunzhang@njust.edu.cn}}

\maketitle


\begin{abstract}
APR (Automated Program Repair) aims to automatically locate program defects, generate patches and validate the repairs. Existing techniques for APR are often combined with LLMs (Large Language Models), which leverages the code-related knowledge of LLMs to improve repair effectiveness. Current LLM-based APR methods typically utilize test cases only during the inference stage, adopting an iterative approach that performs repair first and validates it through test execution afterward. This conventional paradigm neglects two important aspects: the potential contribution of test cases in the training phase, and the possibility of leveraging testing prior to repair. To address this, we propose Repair-R1, which introduces test cases into the model's training phase and shifts test generation to precede repair. The model is required to first generate discriminative test cases that can distinguish defective behaviors, and then perform repair based on these tests. This enables the model to better locate defects and understand the underlying causes of defects, thereby improving repair effectiveness. We implement Repair-R1 with three different backbone models, using RL (reinforcement learning) to co-optimize test generation and bug repair. Experimental results on four widely adopted benchmarks demonstrate the superiority of Repair-R1. Specially, compared to vanilla models, Repair-R1 improves repair success rate by 2.68\% to 48.29\%, test generation success rate by 16.38\% to 53.28\%, and test coverage by 0.78\% to 53.96\%. We publish the code and weights at \href{https://github.com/Tomsawyerhu/APR-RL}{Github} and \href{https://huggingface.co/tomhu/Qwen3-4B-RL-5000-step}{HuggingFace}.
\end{abstract}

\section{Introduction}
APR (Automated Program Repair) aims to automatically locate and fix potential code-related bugs, preventing unexpected behavior under specific inputs, thereby improving the reliability of the program. Since the emergence of LLMs (Large Language Models), they have been widely applied to APR. LLMs learn typical error patterns and repair strategies from existing defect datasets, enabling the generalization of repair capabilities to extensive scenarios.

However, existing LLM-based APR approaches primarily use buggy programs and patches during training, and largely ignore test cases, treating them merely as tools for patch validation.
This can lead to two main problems. First, the model becomes overly dependent on similar bugs. Through pre-training and fine-tuning, the model tends to match syntactically and semantically similar bugs from its parameterized knowledge when encountering new ones, attempting to resolve current issues based on past experience. Although different bugs may appear very similar in text and functionality, their root causes are often quite distinct. As a result, the model may generate superficially similar patches without truly understanding or deeply analyzing the underlying defect. This is akin to relying on rote memorization and doing large volumes of exercises to improve exam performance—an approach that often backfires. The second issue is the underutilization of test data. Test cases are among the most effective tools for identifying bugs, and software projects are often accompanied by a large number of unit tests to ensure code quality. Moreover, during the inference phase, developers typically use error messages to guide the model in performing bug repair. These facts sufficiently highlight the critical role that test cases play in the repair task. Therefore, ignoring test information during the training phase of repair models represents a significant waste. Simply using test cases for test-time scaling is insufficient; instead, test cases should be incorporated into the training phase to enable scaling at the training stage of the repair model.

\begin{figure*}[htb]
  \centering
  \includegraphics[width=\linewidth]{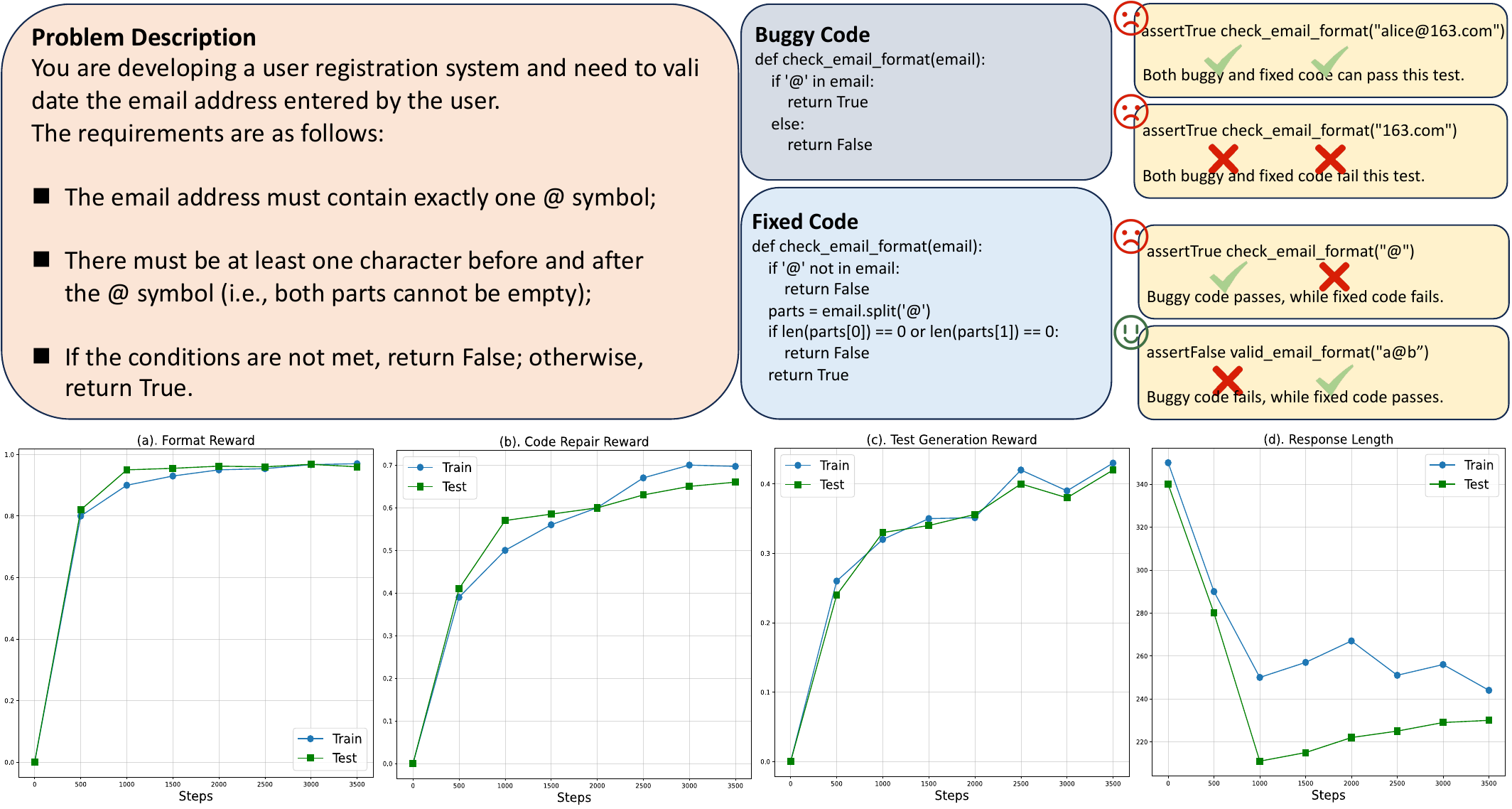}
  \caption{This is a real-world example of bug fixing related to email format validation. To fix the bug, the APR model possibly generates four types of test cases. The first test case passes both the buggy code and the correct code, and thus does not have the ability to expose the bug. The second test case fails in both the buggy and correct code, making it completely incorrect and having no reference value. The third test case passes the buggy code but fails in the correct one — which is exactly the opposite of what is expected. The fourth test case passes the correct code but fails in the buggy code, precisely covering the scenario where the bug is triggered. For the purpose of repair, the fourth type of test case has the capability to distinguish between buggy and correct code, and can help the model understand the bug, thereby improving the success rate of the repair. We expect that during the repair process, the model will generate as many of the fourth-type test cases as possible. To this end, we train the model using RL to jointly optimize test generation and bug repair capabilities. Figures (a)-(d) reflect the simultaneous improvement in test generation and bug repair performance during the model’s (use Qwen2.5-Coder-1.5B-Instruct as an example) training and testing phases.} 
  \label{demo}
\end{figure*}

To address the above issues, we propose Repair-R1, which organically combines the tasks of test generation and APR during the training phase. The core idea of Repair-R1 is illustrated in Figure~\ref{demo}, with the aim of enabling the model to first generate test cases that help locate the bug before generating patches. Specifically, such discriminative test cases have two characteristics: first, they must pass the correct code, ensuring the correctness of the test cases themselves; second, they should fail the buggy code, demonstrating their ability to distinguish between correct and defective implementations. To implement Repair-R1, we design a novel reinforcement learning (RL) framework, as shown in Figure~\ref{overview}.  Each optimization step of Repair-RL is divided into two stages: first, the test generation phase produces discriminative test cases based on the bug, and then the test generation reward is computed according to the effectiveness ratio of the generated test cases. Next, patches are generated based on the given bug and the discriminative test cases, and its correctness is evaluated by running oracle tests, from which the code repair reward is calculated. Finally, the policy model is jointly optimized by combining both rewards. To enhance the effectiveness of the oracle tests, we also perform test augmentation.

Although the correct code is used as the ground-truth patch to validate the correctness of generated test cases, this utilization is implicit. The model does not directly learn the correct patches in a supervised manner, but instead explores the space of possible solutions through self-play. Analogous to solving problems, the supervised learning approach is like a teacher directly revealing the answers, whereas Repair-R1’s training is more like solving the problem on one's own and then being told whether the answer is correct without being shown the details of the solution. Compared to APR models trained via supervised learning, Repair-R1 demonstrates better generalization ability. The result curves in Figure~\ref{demo} also indicate that as the testing capability improves, the repair capability of Repair-R1 is further enhanced.
We summarize our contributions as follows:

(1) We propose Repair-R1, a novel method that utilizes reinforcement learning to train models to generate discriminative tests prior to code repair. This approach helps in locating and understanding bugs, while simultaneously improving the model's abilities in both test generation and bug repair. To the best of our knowledge, Repair-R1 is the first work that employs reinforcement learning to leverage test generation as a guiding mechanism for optimizing the code repair process. 

(2) We validate the effectiveness of Repair-R1 under a configuration involving three models, four benchmarks, and five different methods (a total of 60 experiments). Experimental results show that compared to the untrained model, Repair-R1 improves repair success rate by 2.68\% to 48.29\%, test generation success rate by 16.38\% to 53.28\%, and test coverage by 0.78\% to 53.96\%. Through comparative experiments and ablation studies, we also demonstrate the superiority of Repair-R1 over SFT and single-objective optimization.

(3) Repair-R1 points out a new direction for LLM-based APR, breaking away from the conventional test-driven repair paradigm of "repairing first and testing later," and offers a novel perspective on the repair process.

\section{Background and Related Works}

\subsection{Automated Program Repair}

APR aims to assist developers in localizing and fixing program bugs automatically. 
Traditional APR techniques can be classified as heuristic-based~\citep{le2011genprog,martinez2016astor,yuan2018arja}, constraint-based~\citep{durieux2016dynamoth,martinez2018ultra,mechtaev2016angelix} and template-based~\citep{liu2019tbar,zhang2023gamma}. Although effective, traditional APR approaches are plagued by issues such as high computational cost and limited generalization ability.

Modern APR methods, primarily based on deep learning, have improved upon the shortcomings of previous APR methods. 
As part of learning-based methods, Neural Machine Translation (NMT) techniques have
been extensively studied in recent years, e.g., TENURE~\citep{meng2023template}, Tare~\citep{zhu2023tare},
SelfAPR~\citep{ye2022selfapr}, RewardRepair~\citep{ye2022neural}. They share the same insight that APR can be viewed as an NMT problem that aims to translate buggy code into correct code.
One downside of NMT-based methods is that they relatively sensitive to noise in the training dataset. To improve this, LLM-based methods~\citep{bouzenia2024repairagent,zubair2025use,yang2024cref} leverages the general capabilities of LLMs in code-related tasks, enabling them to fix bugs through zero-shot or few-shot methods, thereby reducing the dependence on high-quality training datasets. 

However, most existing APR models are trained only on buggy programs and patches, ignoring the importance of test information. Test cases are used merely as tools to validate patch correctness during the inference phase. The insufficient utilization of test cases in APR inspires us to incorporate test information into the training phase of APR models. We combine the test generation task with the APR task and leverage RL to jointly optimize the model's ability in both test generation and bug repair. Consequently, the model can understand and fix bugs better by generating discriminative test cases that reveal the root cause of bugs. 

\subsection{Reinforcement Learning}

Proximal Policy Optimization (PPO)~\citep{schulman2017proximalpolicyoptimizationalgorithms} is one of the most popular and representative RL algorithm, which uses an actor-critic setup with clipped updates for stability. Different from PPO, Direct
Preference Optimization (DPO) and its variants ~\citep{rafailov2023direct,liu2020ipo,cen2024value,meng2024simpo} skip the critic and directly optimize from preferences
using closed-form rewards, improving efficiency. Recent efficient Group Relative Policy Optimization (GRPO)~\citep{shao2024deepseekmath} scales well with large-scale reinforcement learning~\citep{guo2025deepseek} by optimizing policies through relative preferences among groups of actions, offering improved stability and performance over traditional methods. Reinforcement learning applied specifically to
coding tasks has also gained traction~\citep{dou2024stepcoder,li2024acecoder}.  

In this paper, we adopt GRPO to optimize Repair-R1 not only for performance considerations. In addition, through mathematical derivation, we reformulate the co-optimization problem as an ELBO (Evidence Lower Bound) maximization problem. We further find that, with appropriate reward design, the optimization objective of GRPO aligns with that of ELBO, making GRPO a well-suited choice for Repair-R1.

\section{Approach}

\begin{figure*}[htb]
  \centering
  \includegraphics[width=\linewidth]{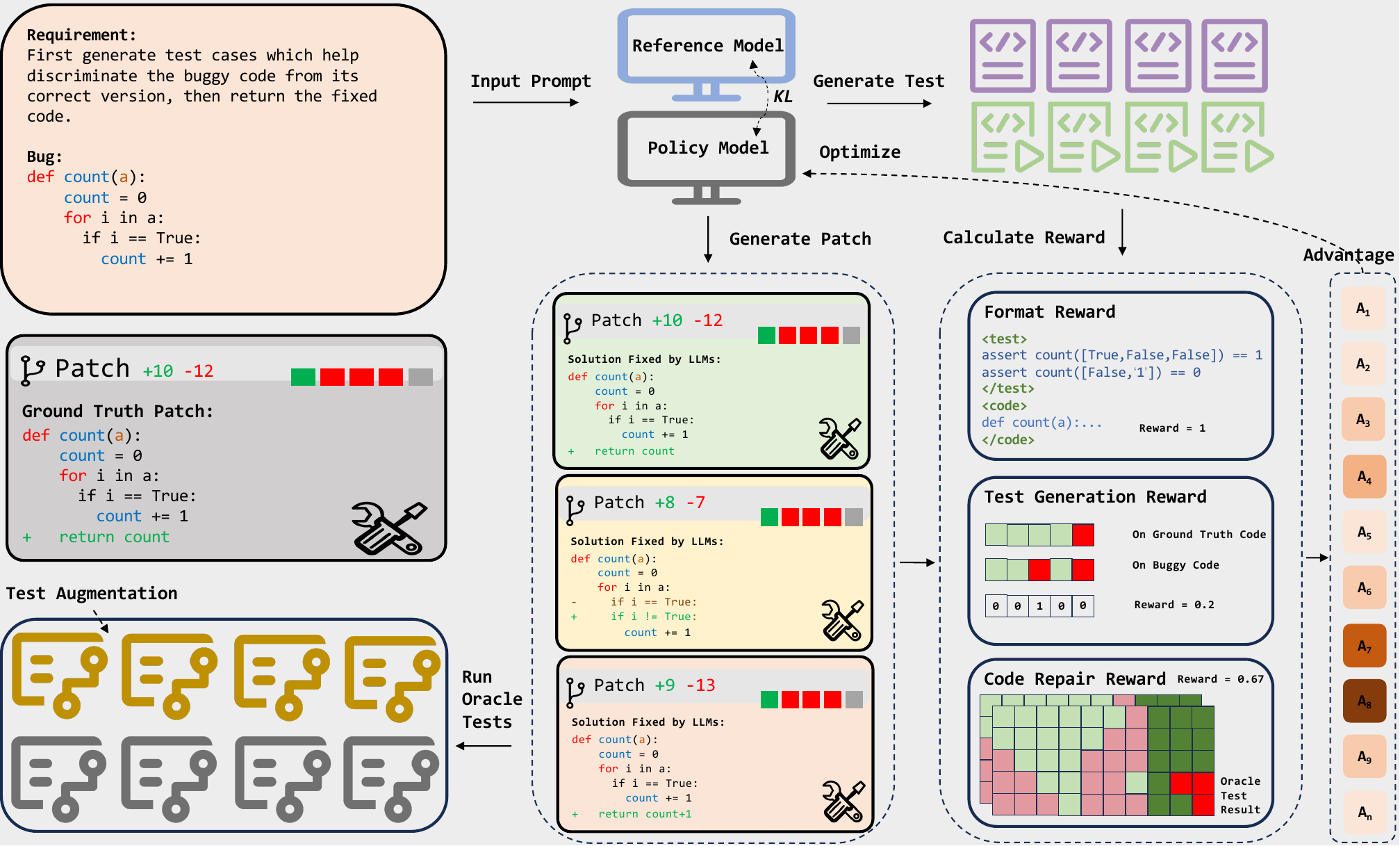}
  \caption{The Optimization Process of Repair-R1} 
  \label{overview}
\end{figure*}

\begin{figure*}[htb]
  \centering
  \includegraphics[width=0.6\linewidth]{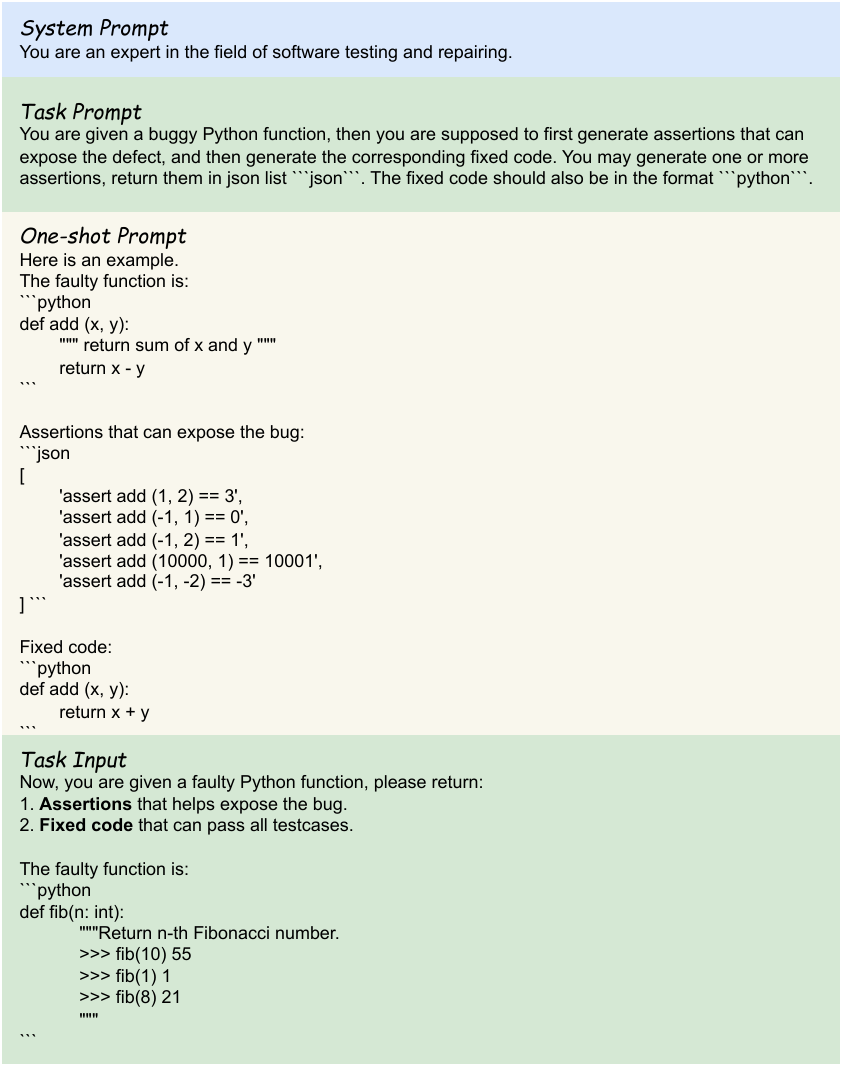}
  \caption{The prompt used for optimizing Repair-R1.} 
  \label{prompt}
\end{figure*}

Figure~\ref{overview} illustrates the overall optimization process of Repair-R1. Before starting the optimization, we copy the original model and refer to it as the reference model, and then perform optimization on the current policy model. At each optimization step, we input the buggy code into the model, and first ask the policy model to generate test cases that can expose the bug. Based on these test cases and the bug, we then ask the policy model to generate corresponding patches. The prompt is given in Figure~\ref{prompt}. We collect the test cases and patches produced by the policy model and evaluate their effectiveness. Specifically, we adopt a reinforcement learning approach to jointly optimize the policy model's test generation capability and code repair capability. Separate rewards are computed for the policy model in terms of test generation and code repair, and a format reward is also introduced to standardize the output format and syntax. Subsequently, advantages are calculated for each sample using the obtained rewards, and the KL divergence between the policy and reference models is computed. The policy model is then updated using both the advantage and the KL divergence. Details on the reinforcement learning algorithm design and the reward computation can be found in Sections ~\ref{reinforcement-learning-algorithm} and ~\ref{reward-modeling}, respectively.

\subsection{Motivation}
Given a bug, the bug repair model will attempt to fix it based on the code-related knowledge it has acquired from pre-trained datasets. The success of repairs may stem from one of two possibilities.

(1) The model accurately identifies the underlying cause of the bug and applies a targeted fix.

(2) The model recovers a similar solution by leveraging memorized patterns from the training data. 

Clearly, the former case is preferable, as it reflects the model's capacity for generalization and its ability to address novel and more complex bugs. Therefore, our motivation is to enhance the model's ability to understand and identify bugs, thereby enabling it to achieve stronger code repair capabilities.

\subsection{Joint Optimization of Repair and Test}
Traditional LLM-based APR is formalized as:

\begin{equation}
\begin{aligned}
\pi(p, b) = \prod_{i=1}^{n} \pi(p_i \mid p_{<i}, b)
\end{aligned}
\end{equation}

Where $\pi$ represents the probability distribution of the repair model, $b$ represents the bug under repair, and $p$ represents a n-token patch output by the model $\pi$. As discussed in the motivation section, we expect the model $\pi$ to comprehend the underlying cause of the bug $b$ before generating the corresponding patch $p$. Thus, we introduce a latent variable $\bm{z}$ to represent the underlying cause of the bug. After the introduction of $\bm{z}$, the code repair task can be further formalized as follows.

\begin{equation}
\begin{aligned}
\pi (\textit{p}, \textit{b}) = \int \pi (\textit{p} \mid \bm{z}, \textit{b}) \cdot \pi(\bm{z} \mid \textit{b}) \, d\bm{z}
\end{aligned}
\end{equation}

Our goal is to optimize the model parameter $\theta$.

\begin{equation}
\begin{aligned}
\theta^* = \max_{\theta} \mathbb{E}_{b \sim D} \left[ \log \pi_\theta(p, b) \right]
\end{aligned}
\end{equation}

Since $ \pi_\theta(p, b) $ involves an intractable integral over the latent variable $ \bm{z} $, direct computation or maximization is not feasible. To address this, we introduce an approximate posterior distribution $ q_\phi(\bm{z} \mid p, b) $ to approximate the true posterior $ \pi_\theta(\bm{z} \mid p, b) $. Based on this approximation, we derive a lower bound of the log-likelihood known as ELBO, which serves as the optimization objective. ELBO is defined as follows.

\begin{equation}
\begin{aligned}
\label{elbo}
\text{ELBO} = \mathbb{E}_{q_\phi(\bm{z} \mid p, b)} \left[ \log \pi_\theta(p \mid \bm{z}, b) \right] - \text{KL}\left(q_\phi(\bm{z} \mid p, b) \,\|\, \pi_\theta(\bm{z} \mid b)\right)
\end{aligned}
\end{equation}

In the field of code repair, as unit tests are commonly employed to evaluate patch correctness, identifying a test case that discriminates between the correct and buggy versions effectively corresponds to pinpointing the underlying bug cause. Therefore, $\pi_\theta(\bm{z} \mid b)$ is converted into a test generation task, which is optimized by both the correctness of generated test cases and their capabilities to discriminate between correct and buggy code. Reward design for test generation is detailed in Section ~\ref{reward-modeling}. Before generating patches, the model is first asked to generate such test cases, and then output patches based on the generated test cases and the original bug. $\mathbb{E}_{q_\phi(\bm{z} \mid p, b)} \left[ \log \pi_\theta(p \mid \bm{z}, b) \right] $ reflects the correctness of the generated patches, we optimize this objective by enhancing the pass rate of the patches on the oracle test cases. Meanwhile, the ELBO optimization objective encourages $\text{KL}\left(q_\phi(\bm{z} \mid p, b) \,\|\, \pi_\theta(\bm{z} \mid b)\right)$ to be as small as possible, which ensures that the optimized model parameters remain close to the original ones.

\subsection{Reinforcement Learning Algorithm}
\label{reinforcement-learning-algorithm}

By incorporating the latest model optimization techniques, we find that the GRPO (Group Relative Policy Optimization) algorithm aligns well with our task objectives. GRPO is used by DeepSeek-R1 and offers strong advantages in balancing cost and performance. On one hand, through reward design, the objective function of GRPO can encompass the optimization goals within the ELBO Formula ~\ref{elbo} , including minimizing KL divergence and maximizing patch correctness. On the other hand, GRPO is well-suited for unsupervised training, where we rely solely on the test pass rate as the optimization signal. By avoiding the use of any supervised learning algorithms, we prevent the model from memorizing patches in the training data, leading to improved generalization performance. In this study, we adopt GRPO to optimize the code repair model. GRPO aims to maximize the following objective function.

\begin{equation}
\label{eq:grpo}
\begin{aligned}
\mathcal{J}_{\text{GRPO}}(\theta)
&= 
\mathbb{E}_{x\sim D,\;y_{1:G}\sim\pi_{\text{old}}(\cdot\!\mid x;\mathcal{C})}
\Biggl[
    \frac{1}{G}\sum_{i=1}^{G}\frac{1}{|y_i|}\sum_{t=1}^{|y_i|}
    \min\!\Bigl(
        \frac{\pi_\theta\!\bigl(y_{i,t}\mid x,y_{i,<t};\mathcal{C}\bigr)}
             {\pi_{\text{ref}}\!\bigl(y_{i,t}\mid x,y_{i,<t};\mathcal{C}\bigr)}
        \,\hat A_{i,t}, \\[4pt]
&\qquad\qquad
        \operatorname{clip}\!\Bigl(
            \frac{\pi_\theta\!\bigl(y_{i,t}\mid x,y_{i,<t};\mathcal{C}\bigr)}
                 {\pi_{\text{ref}}\!\bigl(y_{i,t}\mid x,y_{i,<t};\mathcal{C}\bigr)},
            1-\epsilon,\;1+\epsilon
        \Bigr)\hat A_{i,t}
    \Bigr)
\Biggr]
\;-\;
\beta\,\mathbb{D}_{\mathrm{KL}}\!\bigl[\pi_\theta \,\|\, \pi_{\text{ref}}\bigr],
\end{aligned}
\end{equation}

Where $\epsilon$ and $\beta$ are hyper-parameters, $\hat A_{i,t}$ is the advantage, computed from the relative rewards of responses within each group, and $y_{i,t}$ is $t_{th}$ token generated by the policy model. GRPO incorporates the KL divergence term $\mathbb{D}_{\mathrm{KL}}$ between the policy model and the reference model directly into the loss. During each step of training, GRPO first samples a group of responses from the policy model, then calculate rewards, advantages and KL divergence term for further optimization.

\subsection{Reward Modeling} 
\label{reward-modeling}
In Repair-R1, we employ rule-based rewards instead of using a reward model for the following reasons. First, rule-based rewards have been proven to be effective on code-related tasks. Since the correctness of a patch is generally evaluated by oracle test cases, pass rate can naturally serve as a reliable and interpretable signal for measuring patch quality. Second, rule-based rewards are more straightforward to design and debug compared to learned reward models, which often require large amounts of annotated data and extensive training.  

Specifically, we design three types of rule-based rewards: format reward, code repair reward, and test generation reward. The format reward ensures that the generated outputs are syntactically correct and follow the desired formatting conventions.
The code repair reward encourages the model to produce successful and semantically accurate code repairs. The test generation reward promotes the generation of valid and discriminative test cases that are useful for evaluating the quality of patches. Equal weights are assigned to all three types of rewards.

\textbf{Format Reward.} 
The format reward is designed to enforce a specific output structure for the model. Specifically, we introduce text-based format reward to encourage the model to generate responses in the $<test>...</test><patch>...</patch>$ format, ensuring structured outputs for both code repair and test generation. In contrast to DeepSeek-R1, we do not require the use of $<think>...</think>$, as we observe that excessive thinking steps can lead to performance degradation in code-related tasks. In addition to the text-based reward, we incorporate a syntax-based format reward to ensure that the model's outputs adhere to syntactically valid formats. For code, we check syntactic correctness, while for test cases, it requires a structured JSON list with multiple supporting specifications (e.g., assertions, input/output-style). The format reward is formalized as:

\begin{equation}
\begin{aligned}
R_{\textit{f}} (a) = 
\underbrace{\alpha \cdot  \mathcal{T}(a)}_{\text{Text-based reward}} + \underbrace{\beta \cdot \left[ \mathcal{C}(a_{\text{code}}) + \mathcal{T}(a_{\text{test}}) \right]}_{\text{Syntax-based reward}}
\end{aligned}
\end{equation}

Where $a$ represents the response of the model, $a_{code}$ represents the patch extracted from $a$, $a_{test}$ represents the test cases extracted from $a$, $\mathcal{T}$ represents the result of format validation, and $\mathcal{C}$ represents the result of code compiling.

\textbf{Code Repair Reward.} The code repair reward assesses the effectiveness of the repair process and reflects both the correctness and quality of the generated patches. Rather than treating patch correctness as a binary outcome (correct/incorrect), we model it as a continuous value in the range [0, 1], determined by the pass rate of oracle test cases. A pass rate of 100\% implies full correctness, whereas a 0\% pass rate indicates complete failure.

Although the original datasets typically contain a sufficient number of oracle test cases (often ten or more), to improve the robustness of test-based evaluation, we conduct test case augmentation. The augmented test cases are validated using manually written reference patches to ensure their correctness and exclude invalid ones. The code repair reward is formalized as:

\begin{equation}
\begin{aligned}
R_{\textit{r}}(a) = \frac{1}{|T|} \sum_{t \in T} f(t, a_{\text{code}}) 
\end{aligned}
\end{equation}

Where $a$ represents the response of the model, $a_{code}$ represents the patch extracted from $a$, $T$ represent the augmented set of test cases. For each test case $t$ in $T$, we evaluate $a_{code}$ on $t$, and $f(t, a_{\text{code}}) \in \{0, 1\} $ depends on whether $a_{code}$ passes $t$. Then, the total pass rate of $T$ is calculated as the code repair reward $R_r$.

\textbf{Test Generation Reward.} First, we categorize the generated test cases into two categories: valid and invalid, based on their execution outcomes on both the buggy and fixed versions of the code.  A valid test case satisfies two conditions. First, it is correct, with input-output behavior that matches the expected specification. Second, it is discriminative, meaning it can distinguish between the buggy and correct versions of the code, thereby helping the model locate the defect. Formally, the validity of a generated test case is defined as:

\begin{equation}
\begin{aligned}
V_{\textit{t}} = f_t(t, G) \cdot (1-f_t(t, B))
\end{aligned}
\end{equation}

Where the validity $V_t$ of a test $t$ is equal to 1  if and only if the ground-truth code $G$ passes  $t$ ($f_t(t, G)$ = 1) and the buggy code $B$ fails $t$ ($f_t(t, B)$ = 0). n all other cases,  $V_t$ is set to 0.
Building upon this, the test generation reward further assesses the model's capability of generating multiple test cases and is formally defined as:

\begin{equation}
\begin{aligned}
R_{\textit{t}} = \frac{1}{n} \sum_{i=1}^n V_i = \frac{1}{n}(\mathbf{1}
 - \mathbf{F}) \mathbf{P}^\top
\end{aligned}
\end{equation}

Where vector $ \mathbf{P} = [f(t_1, G), f(t_2, G), \dots, f(t_n, G)] $ indicates whether each test case passes on the ground-truth code $ G $, vector $ \mathbf{F} = [f(t_1, B), f(t_2, B), \dots, f(t_n, B)] $ indicates whether each test case passes on the buggy code $ B $, and $n$ is the number of generated tests. If no tests is generated and $n = 0$, $R_{\textit{t}}$ is set to 0.

\section{Experimental Setup}
\subsection{Dataset Construction}
Dataset construction involves four main steps: collecting normal samples, generating defective variants, performing defect validation and filtering, and train-test split.

\textbf{Normal Samples Collection.} We select four widely used benchmarks for code generation, including HumanEval, MBPP, CodeForces, CodeContests. For HumanEval and MBPP, we included all available samples. In the case of CodeContests and CodeForces, we execute the ground truth solutions on the oracle tests and filter out samples with a runtime of less than 3 seconds. 

\textbf{Defective Variants Generation.} For all collected samples, we use GPT-4o as the mutation model and require it to generate at least 10 defective versions for each sample.

\textbf{Defect Validation and Filtering.} To ensure the validity and distinctiveness of the generated mutants, we first run the oracle test on each mutant to verify that the defect could be detected (i.e., killed by the test). Based on the results, we further remove semantically redundant mutants to maintain the quality and diversity of the dataset.

\textbf{Train-test Split.} Finally, the filtered defect dataset was partitioned into training and test sets at a 4:1 ratio, with care taken to ensure that no original sample appeared in both sets, thereby preventing data contamination. Specifically, the training set contains 5,421 samples, and the test set contains 1,358 samples.

\subsection{Evaluation Metrics}
We evaluate Repair-R1 on the following three metrics:
\begin{itemize}
    \item \textbf{Bugfix}: Bug fix rate.
    \item \textbf{Test}: Success rate of generating test cases that can both pass the ground truth patch and fail the buggy code.
    \item \textbf{Tcov}: The proportion of bugs covered by at least one test that can pass the ground truth patch and fail the buggy code.
\end{itemize}

\section{Experiment Results}
As shown in Table~\ref{tab:qwen_main}, we evaluate three Qwen models with varying sizes and architectures across four defect benchmarks. For each benchmark, we conduct five experimental configurations: (1) using the vanilla model without any adaptation, marked as Vanilla; (2) fine-tuning on defect-repair pairs, marked as SFT; (3) applying reinforcement learning solely on the repair task, marked as RL-Repair; (4) applying reinforcement learning solely on the test generation task, marked as RL-Test; and (5) employing collaborative reinforcement learning on both tasks in a joint manner, marked as RL-Both. In the following sections, we present an analysis of the experimental results.
\begin{table*}[htbp]
\centering
\caption{The performance of Repair-R1 (RL-Both) is assessed across four widely-used benchmarks, including HumanEval, MBPP, CodeForces and CodeContests.}
\label{tab:qwen_main}
\resizebox{\textwidth}{!}{%
\begin{tabular}{l|ccc|ccc|ccc|ccc}
\specialrule{1.2pt}{0pt}{0pt}
\multirow{2}{*}{\textbf{Model}} &
  \multicolumn{3}{c|}{\textbf{HumanEval (112 bugs)}} &
  \multicolumn{3}{c|}{\textbf{MBPP (257 bugs)}} &
  \multicolumn{3}{c|}{\textbf{CodeForces (585 bugs)}} &
  \multicolumn{3}{c}{\textbf{CodeContests (404 bugs)}}\\
& \textbf{Bugfix} & \textbf{Test} & \textbf{Tcov} &
  \textbf{Bugfix} & \textbf{Test} & \textbf{Tcov} &
  \textbf{Bugfix} & \textbf{Test} & \textbf{Tcov} &
  \textbf{Bugfix} & \textbf{Test} & \textbf{Tcov}\\
\hline\hline
Qwen2.5-Coder-1.5B-Instruct (Vanilla)      & 33.04\% & 17.34\% & 37.50\%  & 23.74\% & 9.58\% & 23.74\%  & 4.79\% & 7.99\% & 14.36\%  & 5.45\% & 1.67\% & 2.72\%  \\
Qwen2.5-Coder-1.5B-Instruct (SFT)  & 23.21\% & 3.93\% & 8.04\%  & 9.34\% & 2.09\% & 5.84\%  & 16.75\% & 7.08\% & 9.57\%  & 24.01\% & 1.69\% & 2.48\%  \\
Qwen2.5-Coder-1.5B-Instruct (RL-Test)  & 33.93\% & 48.23\% & 73.21\%  & 28.40\% & 28.99\% & 58.37\%  & 3.25\% & 39.57\% & 51.45\%  & 3.71\% & 50.84\% & 52.48\%  \\
Qwen2.5-Coder-1.5B-Instruct (RL-Repair)  & 75.00\% & 37.54\% & 52.71\%  & 62.65\% & 25.48\% & 32.37\%  & 44.10\% & 31.59\% & 38.72\%  & 41.83\% & 35.85\% & 45.43\%  \\
\textbf{Qwen2.5-Coder-1.5B-Instruct (RL-Both)}  & \textbf{81.25\%} & \textbf{48.07\%} & \textbf{71.43\%}  & \textbf{66.15\%} & \textbf{29.00\%} & \textbf{43.58\%}  & \textbf{46.67\%} & \textbf{40.76\%} & \textbf{50.09\%}  & \textbf{39.85\%} & \textbf{54.95\%} & \textbf{56.68\%}  \\

Qwen2.5-Coder-3B-Instruct (Vanilla)      & 51.79\% & 24.45\% & 56.25\%  & 38.91\% & 18.54\% & 44.36\%  & 10.26\% & 9.89\% & 29.91\%  & 13.12\% & 5.64\% & 16.09\%  \\
Qwen2.5-Coder-3B-Instruct (SFT) & 19.53\% & 6.43\% & 14.06\%  & 22.57\% & 6.68\% & 13.54\%  & 29.02\% & 6.22\% & 15.03\%  & 33.05\% & 6.10\% & 12.53\%  \\
Qwen2.5-Coder-3B-Instruct (RL-Test) & 56.25\% & 48.59\% & 59.82\%  & 46.30\% & 40.70\% & 46.69\%  & 14.70\% & 40.77\% & 61.54\%  & 12.87\% & 55.24\% & 58.42\%  \\
Qwen2.5-Coder-3B-Instruct (RL-Code) & 79.46\% & 26.74\% & 58.93\%  & 68.87\% & 20.04\% & 48.25\%  & 52.31\% & 16.68\% & 41.37\%  & 47.77\% & 8.71\% & 23.02\%  \\
\textbf{Qwen2.5-Coder-3B-Instruct (RL-Both)}   & \textbf{85.71\%} & \textbf{50.10\%} & \textbf{60.71\%}  & \textbf{69.65\%} & \textbf{36.58\%} & \textbf{45.14\%}  & \textbf{53.85\%} & \textbf{41.74\%} & \textbf{63.76\%}  & \textbf{50.00\%} & \textbf{55.22\%} & \textbf{59.16\%}  \\

Qwen3-4B (Vanilla)       & 83.93\% & 40.00\% & 81.25\%  & 48.64\% & 26.18\% & 53.31\%  & 35.21\% & 27.52\% & 57.26\%  & 34.16\% & 11.93\% & 31.44\%  \\
Qwen3-4B (SFT)       & 66.07\% & 8.22\% & 16.96\%  & 67.70\% & 12.69\% & 28.40\%  & 54.70\% & 4.00\% & 5.47\%  & 45.30\% & 1.79\% & 3.47\%  \\
Qwen3-4B (RL-Test)      & 71.43\% & 57.70\% & 77.68\%  & 58.37\% & 50.11\% & 70.04\%  & 31.28\% & 44.55\% & 68.38\%  & 30.94\% & 56.86\% & 62.38\%  \\
Qwen3-4B (RL-Code)       & 84.82\% & 40.50\% & 86.61\%  & 70.82\% & 26.03\% & 61.87\%  & 51.79\% & 30.90\% & 64.62\%  & 51.49\% & 14.99\% & 40.10\%  \\
\textbf{Qwen3-4B (RL-Both)}       & \textbf{86.61\%} & \textbf{59.49\%} & \textbf{73.21\%}  & \textbf{71.98\%} & \textbf{55.27\%} & \textbf{73.15\%}  & \textbf{54.19\%} & \textbf{43.90\%} & \textbf{66.67\%}  & \textbf{51.73\%} & \textbf{58.11\%} & \textbf{61.39\%}  \\
\specialrule{1.2pt}{0pt}{0pt}
\end{tabular}
}
\end{table*}

\subsection{Test Generation Helps Repair Better}

Through RL training, the repair capability of Repair-R1 is significantly improved compared to the vanilla model, and the test generation ability is also enhanced accordingly. As shown in Table ~\ref{tab:qwen_main}, all three models demonstrate consistent improvements in both repair and test generation across the four benchmarks. The repair success rate increases by 2.68\% to 48.29\%, the test generation success rate improves by 16.38\% to 53.28\%, and the test coverage is enhanced by 0.78\% to 53.96\%.

\begin{figure*}[htb]
  \centering
  \includegraphics[width=\linewidth]{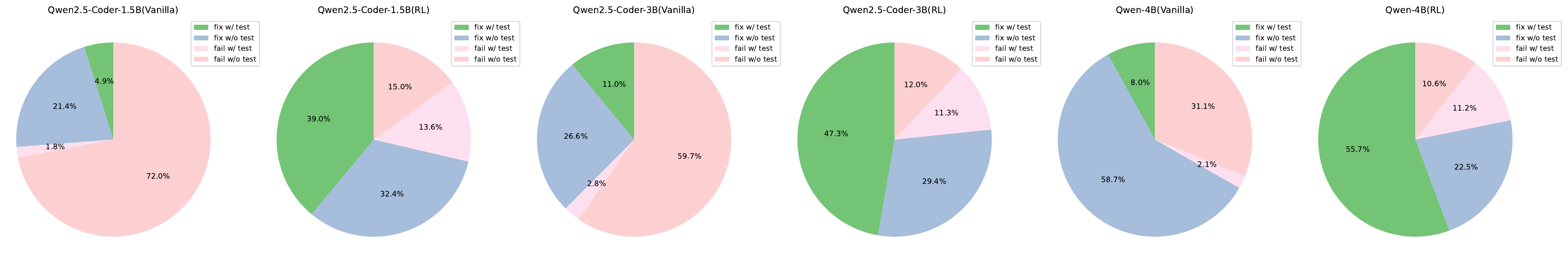}
  \caption{A comparison of test generation and repair correlation before and after RL} 
  \label{pie_chart}
\end{figure*}

To further explore whether enhancements in test generation can facilitate improvements in repair capability, we classify the combinations of repair outcomes and test generation results into four categories: 

(1) Successful repair with effective test case generation, marked as \textbf{fix w/ test}.

(2) Successful repair without effective test case generation, marked as \textbf{fix w/o test}.

(3) Failed repair with effective test case generation, marked as \textbf{fail w/ test}.

(4) Failed repair without effective test case generation, marked as \textbf{fail w/o test}.

As shown in Figure ~\ref{pie_chart}, we compare the distribution of four types of samples across the three models before and after RL training. We observe that for the vanilla models, the ability to generate tests is highly limited, which in turn constrains the effectiveness of repair. Most successful repairs are accompanied by failed test cases and generally exhibit low success rates, indicating that the model merely outputs syntactically correct code without understanding the underlying cause of the defect. In contrast, after RL training, both test generation and repair capabilities are significantly improved. Successful repairs are often associated with correctly generated test cases, suggesting that the model can now perform more effective repairs based on a better understanding of the defect location and its root cause.

\subsection{Repair-R1 Outperforms SFT on Imbalanced Datasets}
To broadly evaluate the generalization capability of Repair-R1 across different types of bugs, we pay special attention to the type of samples when selecting benchmarks. Specifically, HumanEval and MBPP consist of function-level bugs without entry points, while CodeContests and CodeForces use standard input and output formats. Moreover, the dataset exhibits an imbalanced distribution, with HumanEval and MBPP contributing only a small portion (approximately one-third) of the total samples. 

The results in Table~\ref{tab:qwen_main} indicate that the disparities in data types and the imbalance in distribution have notably impacted the effectiveness of SFT. For the more prevalent benchmarks, CodeForces and CodeContests, the model shows improved repair performance after SFT. Specifically, Qwen2.5-Coder-1.5B-Instruct improves the repair success rate by 11.96\% and 18.56\%, Qwen2.5-Coder-3B-Instruct by 18.76\% and 19.93\%, and Qwen-4B by 19.49\% and 11.14\% on CodeForces and CodeContests, respectively.
However, for the less-represented benchmarks, HumanEval and MBPP, the model exhibits a forgetting effect after fine-tuning, with a significant drop in repair performance compared to before training. 
Specifically, Qwen2.5-Coder-1.5B-Instruct decrease repair success rate by 9.83\% and 14.4\% and Qwen2.5-Coder-3B-Instruct by 32.26\% and 16.34\% on HumanEval and MBPP, respectively.

In contrast, RL avoids this issue. The model trained with RL demonstrates consistently improved repair performance across all benchmarks. The improvement in repair success rate across the four benchmarks ranges from 2.68\% to 48.29\% for the three models.
By comparing SFT and RL, we find that SFT merely fits the data distribution during training. Since pre-training tasks typically do not involve code repair, significant parameter updates occur when the model is fine-tuned on defect datasets, leading it to overfit to the dominant distributions (e.g., CodeContests and CodeForces), while suffering from forgetting on underrepresented datasets (e.g., HumanEval). In contrast, RL builds upon the model’s foundational coding capabilities and incrementally learns repair patterns, thereby enhancing repair performance without causing catastrophic forgetting.
 
\subsection{Ablation Study}

We conduct an ablation study on the tasks of test generation and code repair. In the RL training process, we retain only the Code Repair Reward or the Test Generation Reward separately, and evaluate the performance of the trained model on both test generation and code repair tasks.

As illustrated in Table~\ref{tab:qwen_main}, training the model exclusively on test generation appears to result in a marginal improvement in repair capability. This observation is supported by certain results. For instance, Qwen2.5-Coder-3B-Instruct (Test) achieves an increase in repair success rates of 4.46\%, 7.39\%, and 4.44\% on HumanEval, MBPP, and CodeForces, respectively, compared to the vanilla model. Nevertheless, this improvement is not stable, with Qwen-4B (Test) exhibiting a decrease in repair effectiveness on HumanEval, CodeForces, and CodeContests.

On the other hand, models trained solely on the repair task also show improvements in test generation. This trend is stable, with all three models achieving varying degrees of improvement in test coverage and test effectiveness across four benchmarks. 

Building upon these findings, we compare the two ablation models with Repair-R1 and observe that Repair-R1 demonstrates the most favorable performance in terms of both repair and test generation. Specifically, compared to RL-Repair, Repair-R1 achieves a repair success rate improvement ranging from 0.24\% to 6.25\%, with improvements observed in 11 out of 12 comparison settings. Regarding test generation, Repair-R1 achieves a comparable level of test coverage as RL-Test, while demonstrating a higher success rate in test generation.

\subsection{Test-time Scaling Performance of Repair-R1}
\begin{figure}[htbp]    

  \centering            
  \subfloat[HumanEval]   
  {
      \label{scale_humaneval}\includegraphics[width=0.24\textwidth]{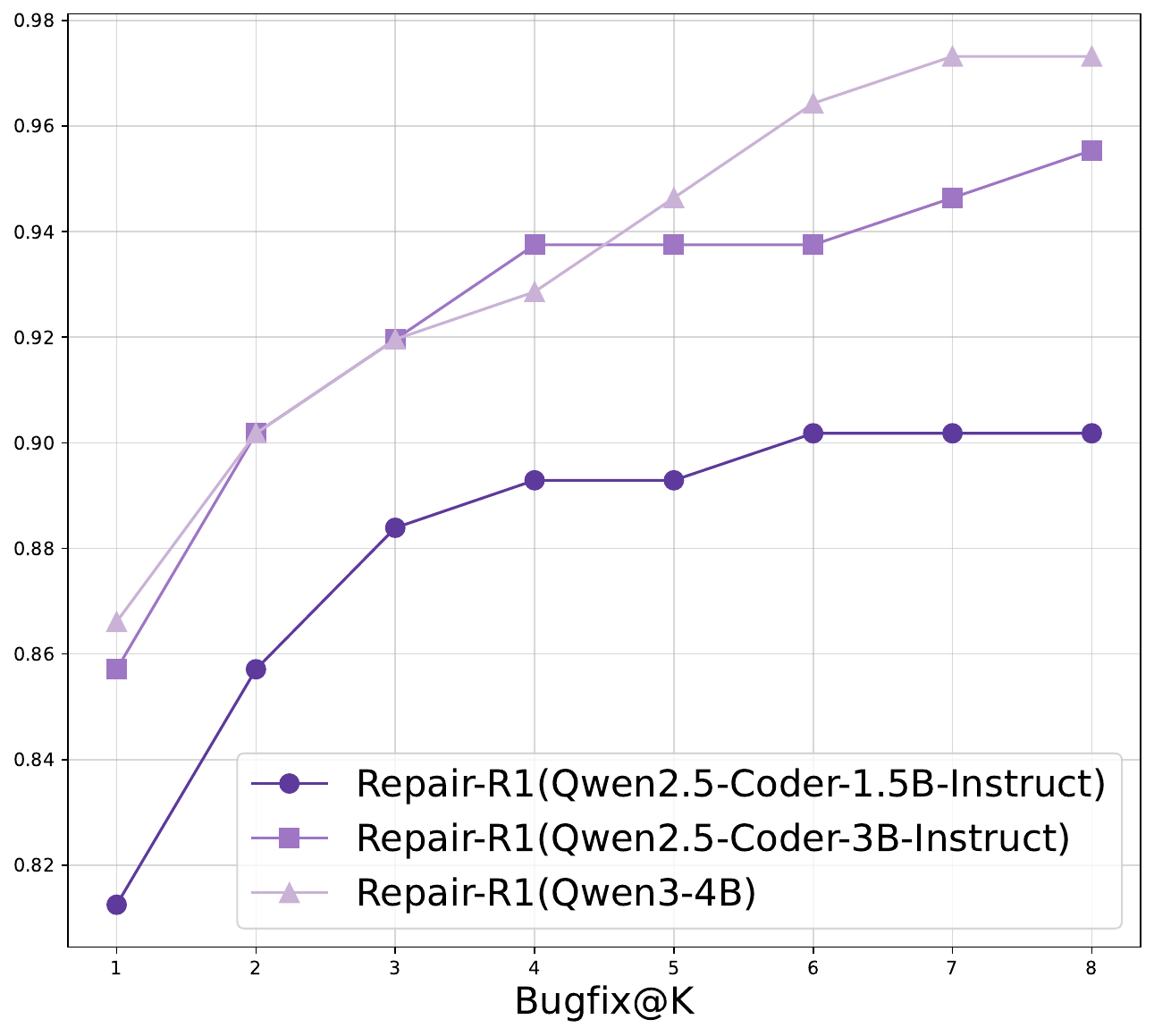}
  }
  \subfloat[MBPP]   
  {
      \label{scale_mbpp}\includegraphics[width=0.24\textwidth]{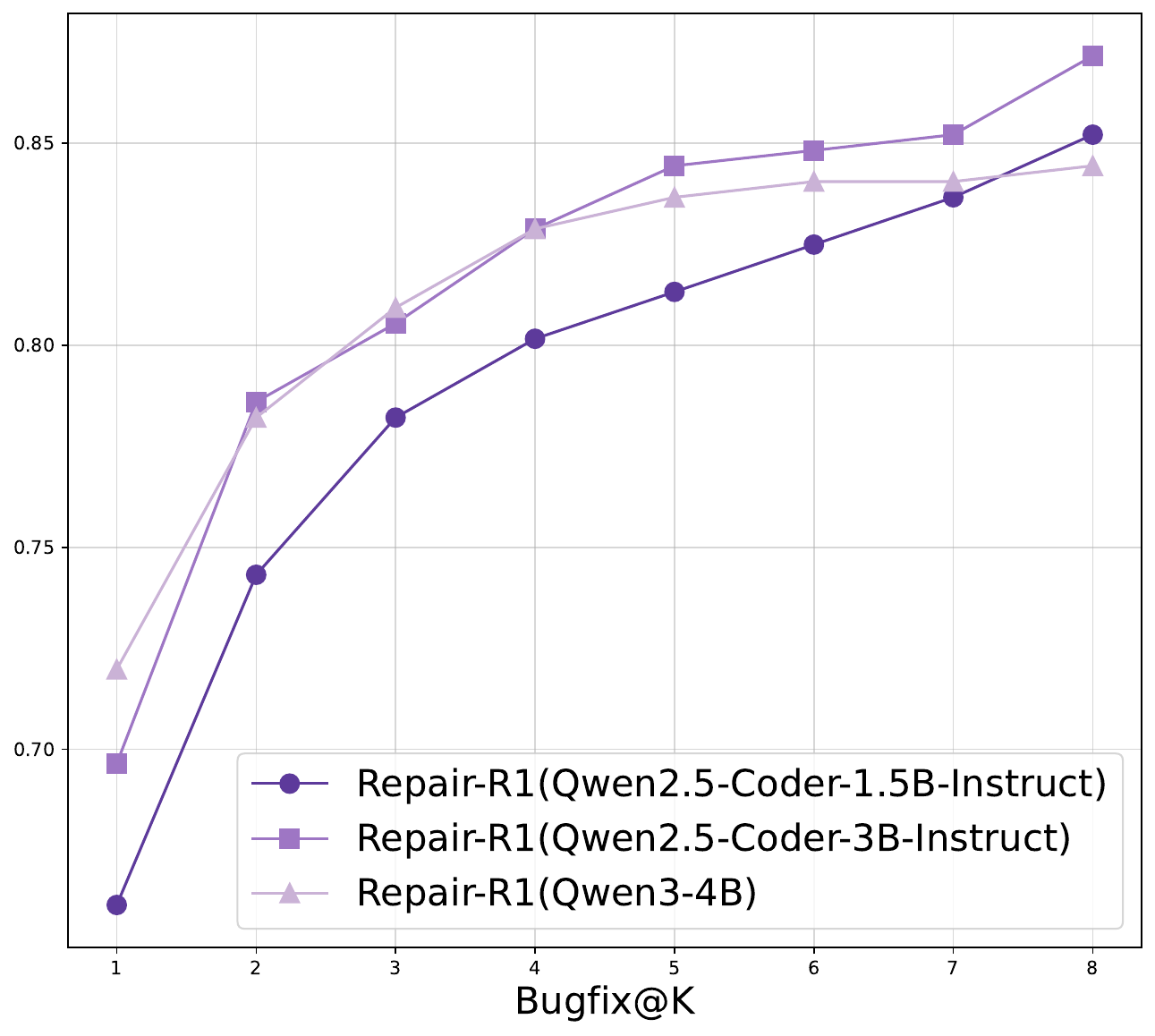}
  }
  \subfloat[CodeForces]   
  {
      \label{scale_codeforces}\includegraphics[width=0.24\textwidth]{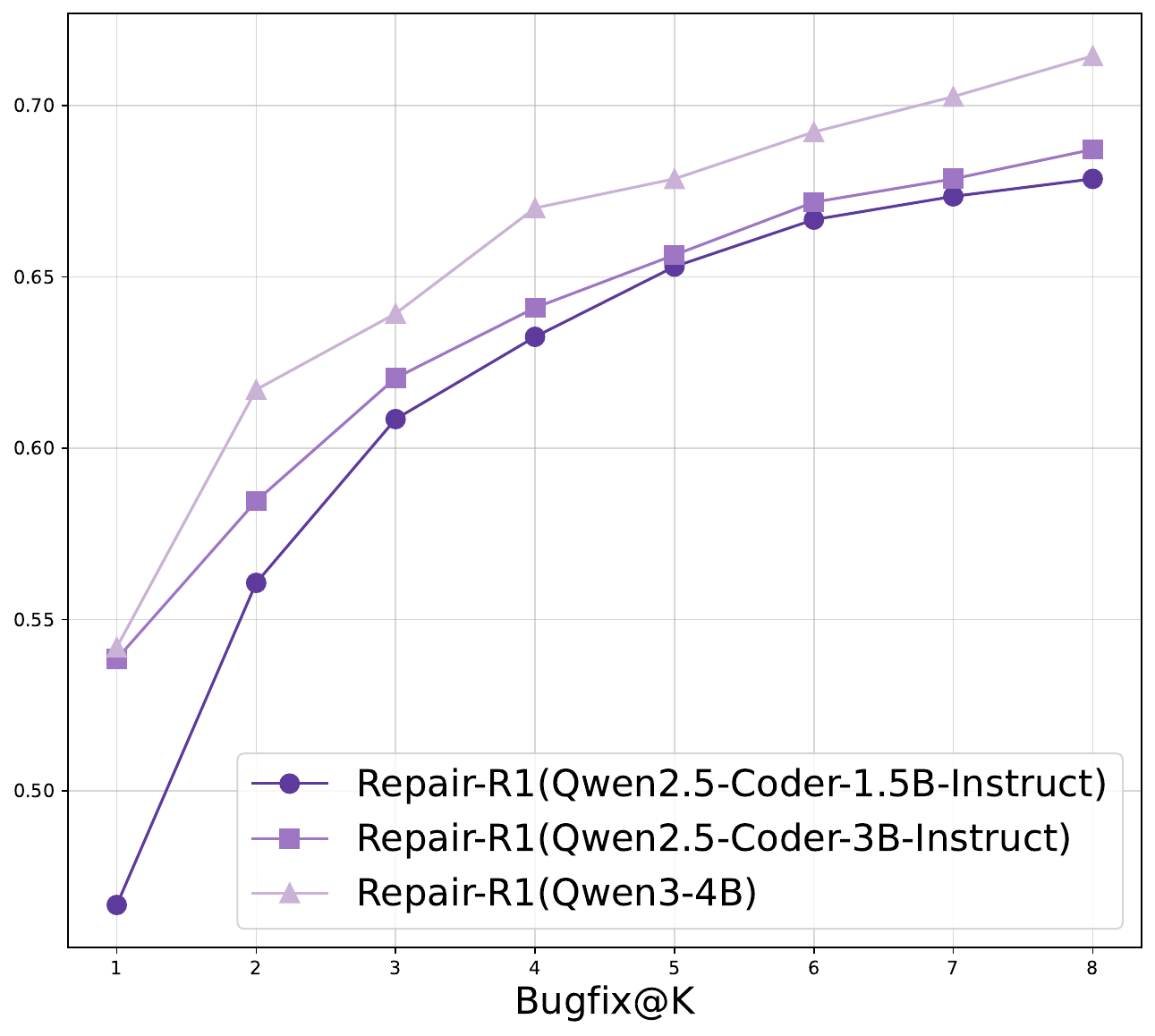}
  }
  \subfloat[CodeContests]   
  {
    \label{scale_codecontests}\includegraphics[width=0.24\textwidth]{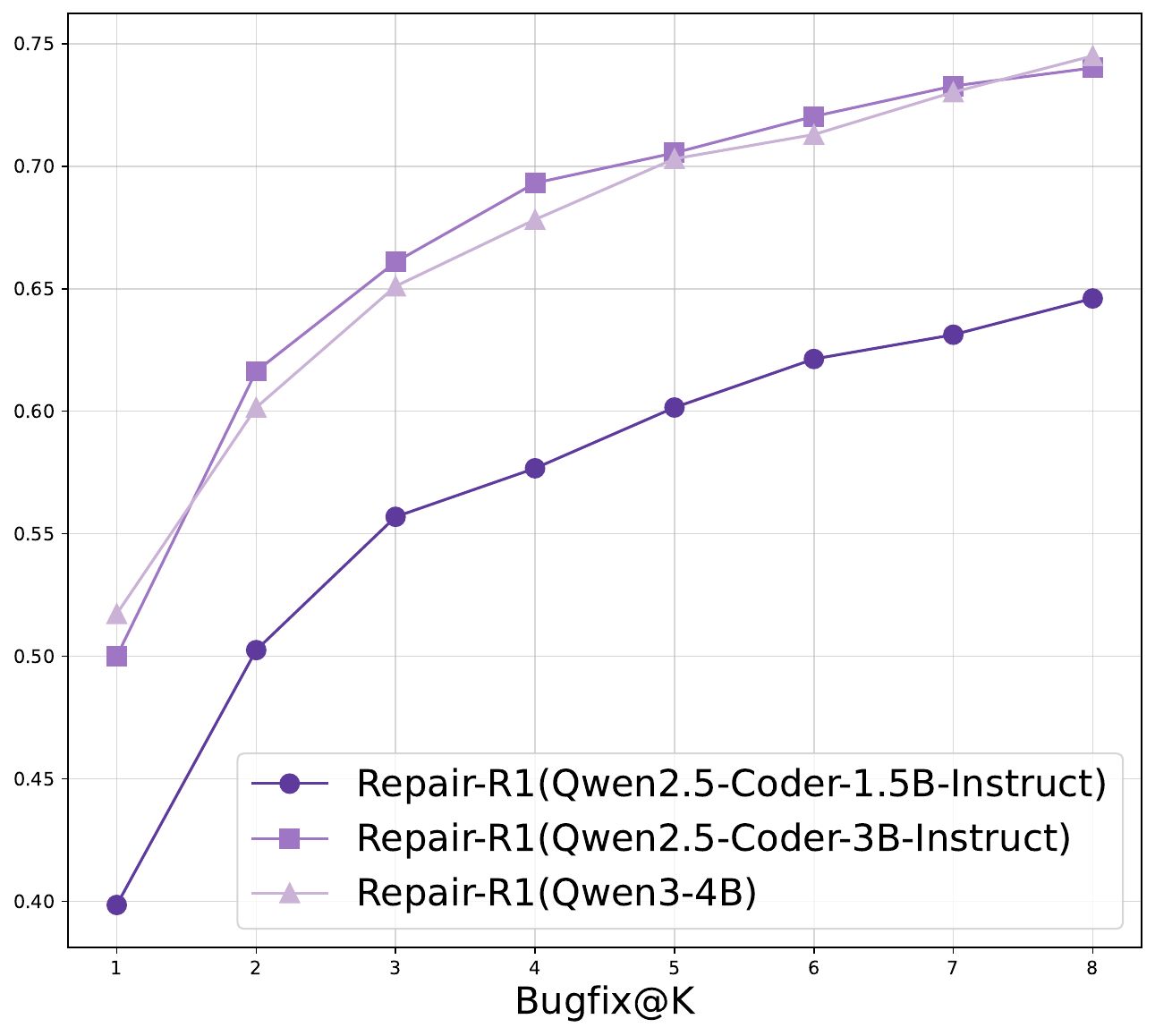}
  }

  \caption{Bugfix@K performance of Repair-R1 on different benchmarks.}
  \label{scale_results}
  
\end{figure}

We analyze the test-time scaling ability of Repair-R1 across different benchmarks and models.
As shown in Figure~\ref{scale_results}, we present the repair success rate of Repair-R1 under different sampling configurations, with sample sizes varying from 1 to 8.

In general, Qwen-4B exhibits the best scaling capability among the three models, achieving superior performance on HumanEval and CodeForces, comparable results with Qwen2.5-Coder-3B-Instruct on CodeContests, and slightly lower performance on MBPP. We believe this is not entirely due to the difference in parameter scale. We observe that when the sampling size is less than or equal to 4, Qwen-4B performs slightly worse than Qwen2.5-Coder-3B-Instruct overall, as the latter is a code-specific language model (CodeLM) and thus has a direct advantage on code-related tasks. However, as the sampling size increases, the repair performance of Qwen-4B gradually surpasses that of Qwen2.5-Coder-3B-Instruct. This can be attributed to the fact that Qwen-4B is a reasoning model, which benefits more from larger sampling sizes by leveraging diverse reasoning paths to generate higher-quality patches. This trend also suggests that while specialized models like Qwen2.5-Coder-3B-Instruct and Qwen2.5-Coder-1.5B-Instruct may have an edge in low-sampling scenarios due to their task-specific training, general-purpose reasoning models such as Qwen-4B can catch up and even outperform them when given more samples, thanks to their broader knowledge base and stronger reasoning capabilities.

\section{Conclusion}

This paper presents Repair-R1, a method that leverages the generation of discriminative test cases to guide the model in performing bug repair, thus jointly improving its abilities in both test generation and bug repair. We design reward functions for both test generation and code repair, and employ the GRPO algorithm to optimize the model. Experimental results show that compared to vanilla models, Repair-R1 improves repair success rate by 2.68\% to 48.29\%, test generation success rate by 16.38\% to 53.28\%, and test coverage by 0.78\% to 53.96\%. Comparative experiments and ablation studies also demonstrate the superiority of Repair-R1 over SFT and single-objective optimization.

\bibliographystyle{IEEEtran}
\bibliography{reference}
\end{document}